\documentclass[a4paper,11pt]{article}
\usepackage{pos}

\title{Deep Learning-Based Stereoscopic Event Reconstruction for CTAO using CTLearn}
\ShortTitle{Stereoscopic CTAO Event Reconstruction using CTLearn}

\manuallySeparateAuthors
\author*[a]{T. Miener}
\author[b]{, I. Viale}
\author[a]{, B. Lacave}
\author[c]{, and A. Cerviño}

\author[d]{ for the CTAO-LST project}
\affiliation[a]{Département de Physique Nucléaire et Corpusculaire, Université de Genève, Faculté de Sciences,\\
1211 Genève 4, Switzerland}
\affiliation[b]{INFN Sezione di Torino, Via P. Giuria 1, 10125 Torino, Italy}
\affiliation[c]{IPARCOS-UCM, Instituto de Física de Partículas y del Cosmos, and EMFTEL Department, Universidad Complutense de Madrid, Plaza de Ciencias, 1. Ciudad Universitaria, 28040 Madrid, Spain}
\affiliation[d]{\href{https://www.lst1.iac.es}{https://www.lst1.iac.es}; see full author list of the CTAO-LST Project at the end of the document}

\emailAdd{tjark.miener@unige.ch}

\abstract{The Cherenkov Telescope Array Observatory (CTAO), a next-generation ground-based gamma-ray observatory, will be composed of two arrays of multiple imaging atmospheric Cherenkov telescopes (IACTs) located in both the Northern and Southern Hemispheres. Its goal is to enhance the sensitivity of current instruments by a factor of five to ten over an energy range from 20 GeV to over 300 TeV. IACT arrays are used to probe the very-high-energy (VHE) gamma-ray sky, operating by simultaneously observing air showers triggered by the interaction of VHE gamma rays and cosmic rays with the atmosphere. Cherenkov photons produced by these showers create a stereoscopic record of the event. By reconstructing the event using machine learning techniques, the properties of the originating VHE particle—including its type, energy, and incoming direction—can be determined. In this contribution, we present a fully deep-learning-driven approach to reconstruct simulated, stereoscopic IACT events using CTLearn. CTLearn is a package designed for loading and manipulating IACT data and for running deep learning models with pixel-wise camera data as input.}

\ConferenceLogo{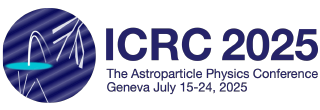}

\FullConference{39th International Cosmic Ray Conference (ICRC2025)\\
15–24 July 2025\\
Geneva, Switzerland\\}

\begin{document}

\maketitle

\section{Introduction}
The Cherenkov Telescope Array Observatory (CTAO)~\cite{2019APh...111...35A} is the next-generation ground-based gamma-ray observatory and will consist of two arrays of tens of imaging atmospheric Cherenkov telescopes (IACTs) to be built in the Northern Hemisphere (La Palma, Canary Island, Spain) and in the Southern Hemisphere (near Cerro Paranal, Chile). CTAO aims to improve the sensitivity of current-generation instruments by a factor of five to ten and provide an energy coverage from 20 GeV to more than 300 TeV thanks to three types of IACTs: the large-sized telescope (LST), the medium-sized telescope (MST), and the small-sized telescope (SST). Among these, the LSTs are designed to be the most sensitive to the lowest gamma-ray energies, from approximately 20 GeV to a few hundred GeV, thanks to their large reflective mirror area (23 meters in diameter) and fast readout electronics optimized for detecting weak Cherenkov light flashes from low-energy showers. The CTAO North site will host four LSTs as its core instruments for low-energy gamma-ray detection. As of now, the first LST (LST-1) has been fully constructed and is undergoing commissioning and scientific validation, while the second LST (LST-4) started its commissioning phase in mid-2025. The remaining two telescopes (LST-3 and LST-2) are currently under construction and expected to be completed in the near future.

In this work, we focus on \texttt{LST1+LST4}, which form the first operational pair capable of performing stereoscopic observations, a key capability that significantly enhances event reconstruction accuracy and background rejection compared to single-telescope observations. These two telescopes represent a milestone for CTAO's early science phase and enable detailed performance studies under operating conditions. Their role is essential for validating analysis pipelines, optimizing observation strategies, and preparing for the full array deployment. The initial quartet of LSTs represents the first phase of CTAO's deployment and is crucial for achieving early science goals, particularly in studies of transient and variable gamma-ray sources such as gamma-ray bursts, active galactic nuclei, and pulsars. Their rapid slewing capability and low-energy sensitivity make them uniquely suited for prompt observations of transient phenomena, reinforcing the CTAO's role as a key instrument in multi-messenger astrophysics.

IACTs collect the Cherenkov light induced by the showers of particles produced when very-high-energy (VHE; approximately above 20 GeV) gamma-rays or charged cosmic rays, the dominant background, enter the atmosphere and focus those Cherenkov photons onto their camera, producing a record of the event. Together with its spatial, temporal, and calorimetric information, the IACT images contain the longitudinal development of the air shower. The driving factors determining the sensitivity of IACTs to astrophysical sources is how well reconstructed the properties (type, energy, and incoming direction) of the primary particle triggering the air shower are. The morphological differences of the gamma-ray and cosmic-ray-initiated showers, translated into their images, make them distinguishable.

The original IACT data analysis classified the triggered events from their camera images by extracting handcrafted features, like the commonly used Hillas parameters~\cite{1985ICRC....3..445H}, and performed parameter-wise selection over the multidimensional space of those parameters. Nowadays, more sophisticated procedures where supervised learning algorithms like Random Forest (RF)~\cite{2008NIMPA.588..424A} or Boosted Decision Trees (BDTs)~\cite{2009APh....31..383O,2011APh....34..858B,2017APh....89....1K} are trained on those features to infer the full-event reconstruction have been incorporated in the analysis workflow, as a result of the improvement in available computational resources and algorithms over the past few decades. Deep convolutional neural networks (CNNs), a particular class of deep learning (DL) algorithms, can be utilized to carry out this task by accessing the information contained in the shower images since these types of algorithms are capable of learning the feature extraction by themselves (representation learning)~\cite{Goodfellow-et-al-2016}.


\section{Data analysis pipeline}

The analysis is carried out with the CTAO-North simulated dataset designed for stereoscopic observations. The dataset was obtained with \texttt{CORSIKA}~\cite{Heck:1998vt} and \texttt{sim\_telarray}~\cite{Bernlohr:2008kv} following the standard procedure of CTAO. The simulations were further reduced from raw to calibrated and cleaned image data using \texttt{ctapipe}~\cite{karl_kosack_2025_15606984}, a prototype low-level data processing pipeline for CTAO. The reduced dataset follows the \texttt{ctapipe} reference implementation of the CTAO data format. We restricted ourselves to simulated data with a Zenith angle of $ 20^{\circ} $ and an Azimuth angle of $ 180^{\circ} $ (South pointing). For the DL training process, simulated diffuse gamma-ray and proton-initiated events are considered, generated within cones of $ 6^{\circ} $ and $ 10^{\circ} $ radius, respectively. The regression models (energy or arrival direction reconstruction) are trained only with the training set of diffuse gamma rays. The performances of the different are evaluated on simulated protons ($ \sim 10^{8} $ events; energy range $ 10~\text{GeV}-100~\text{TeV} $; viewcone $10^{\circ}$) and electrons ($ \sim 10^{8} $ events; energy range $ 5~\text{GeV}-5~\text{TeV} $; viewcone $6^{\circ}$), as well as pointlike gamma rays ($ \sim 10^{7} $ events; energy range $ 5~\text{GeV}-50~\text{TeV} $) assuming a gamma-ray point source in the center of the field of view.

\texttt{CTLearn}\footnote{\href{https://github.com/ctlearn-project/ctlearn}{https://github.com/ctlearn-project/ctlearn}}~\cite{miener_2025_15065761,Nieto:2019ak} was employed for DL-based particle classification and event reconstruction, such as the regression for the energy and the arrival direction. The predictions of \texttt{CTLearn} are written in \texttt{ctapipe}-compatible format, following the CTAO reference data structure, which ensures consistency and interoperability with other components of the Data Processing and Preservation System (DPPS) of CTAO. Following the reconstruction, energy-dependent cut optimization and the construction of Instrument Response Functions (IRFs) were performed using \texttt{ctapipe}, which integrates the \texttt{pyirf}~\cite{maximilian_linhoff_2025_15488321} package to carry out these key post-reconstruction tasks. This seamless workflow highlights the readiness of the DL-based \texttt{CTLearn} pipeline for integration into the broader CTAO data processing ecosystem. To ensure reproducible results with minimal manual management, we used the \texttt{CTLearn-Manager} package, which streamlines the training, evaluation, and data handling processes through an automated and configurable interface.

\section{Methodology}

We adopt the same methodological framework as presented in~\cite{2021arXiv211201828M,Miener:2022xws}, which utilized data from the two MAGIC telescopes. In the presented analysis, we apply the method to simulated data from the \texttt{LST1+LST4}, the first pair of LSTs of the CTAO. Two key aspects define this approach. First, to fully leverage the stereoscopic information provided by the dual-telescope system, we stack the respective images (integrated pixel charges and signal arrival times) from each LST, along the channel dimension before inputting them into a convolutional neural network (CNN). Second, the images are subjected to standard image cleaning procedures, as used in conventional IACT analyses. This cleaning step retains pixels containing significant Cherenkov signal while removing noise pixels that are likely dominated by photons from the night sky background, thus enhancing robustness and ensuring the method remains applicable to real observational data.

For this study, we choose the \textit{CTLearn-stackTRN model}, developed within the \texttt{CTLearn} framework, which is a shallow residual neural network (ResNet)~\cite{2015arXiv151203385H} with 33 layers inspired by the Thin-ResNet (TRN)~\cite{2019arXiv190210107X}. The first initialization layer of the original TRN~\cite{2019arXiv190210107X} is skipped in order to adjust for the specific image input shape of the LST. The residual connections, meaning that the original input is added to the output at each stage, allow exploring much deeper model architecture than plain networks. A dual squeeze-and-excitation attention mechanism~\cite{2017arXiv170901507H} is deployed in each residual block, which helps the network to focus on the most relevant features of the image representation. The images were pre-processed to a Cartesian lattice using the bilinear interpolation routine from the \texttt{ImageMapper} of the \texttt{DL1-Data-Handler}\footnote{\href{https://github.com/cta-observatory/dl1-data-handler}{https://github.com/cta-observatory/dl1-data-handler}}~\cite{tjark_miener_2025_15422957}, a package designed for the data management of machine learning image analysis techniques for IACT data, to apply conventional convolutional layers~\cite{Nieto:2019uj}.

We selected the TRN architecture for our work based on~\cite{2021arXiv211201828M,Miener:2022xws} demonstrating its strong balance between performance and efficiency. The TRN is widely adopted due to its feasibility in training, offering robust results while keeping computational demands moderate. While deeper and more complex models can potentially outperform the TRN in terms of accuracy, they typically require significantly more processing time and computational resources. In our case, training the TRN took less than three hours using two state-of-the-art NVIDIA GH200 GPUs in parallel, and inference required less than 15 minutes on a single GPU of the same kind. We deployed and ran the entire \texttt{CTLearn}-based analysis at the Swiss National Supercomputing Centre (CSCS), which serves as one of the off-site Information and Communications Technology (ICT) facilities for the CTAO.

\section{Results}

This section presents the performance evaluation of the stereoscopic analysis using the \textit{CTLearn-stackTRN model}, applied to simulated data from the first two Large-Sized Telescopes \texttt{LST1+LST4}. The goal is to assess the capability of the DL model to reconstruct and classify events using combined image information from both telescopes. We restrict the analysis to stereo events in which both images contain more than 50 photoelectrons (p.e.) after the cleaning procedure, ensuring sufficient signal quality for accurate reconstruction. We begin with a reconstruction validation, where the reconstructed quantities are directly compared to their true simulated values. This is followed by an energy-dependent cut optimization procedure, which is needed to assess the performance of the reconstruction. We then evaluate the angular and energy resolution as a function of the energy, and finally, we estimate the sensitivity of the system using standard performance metrics. These results serve as a first step toward validating the applicability of \texttt{CTLearn}-based DL methods for stereoscopic LST observations.

\begin{figure}[h]
    \centering
    \begin{overpic}[width=4.9cm]{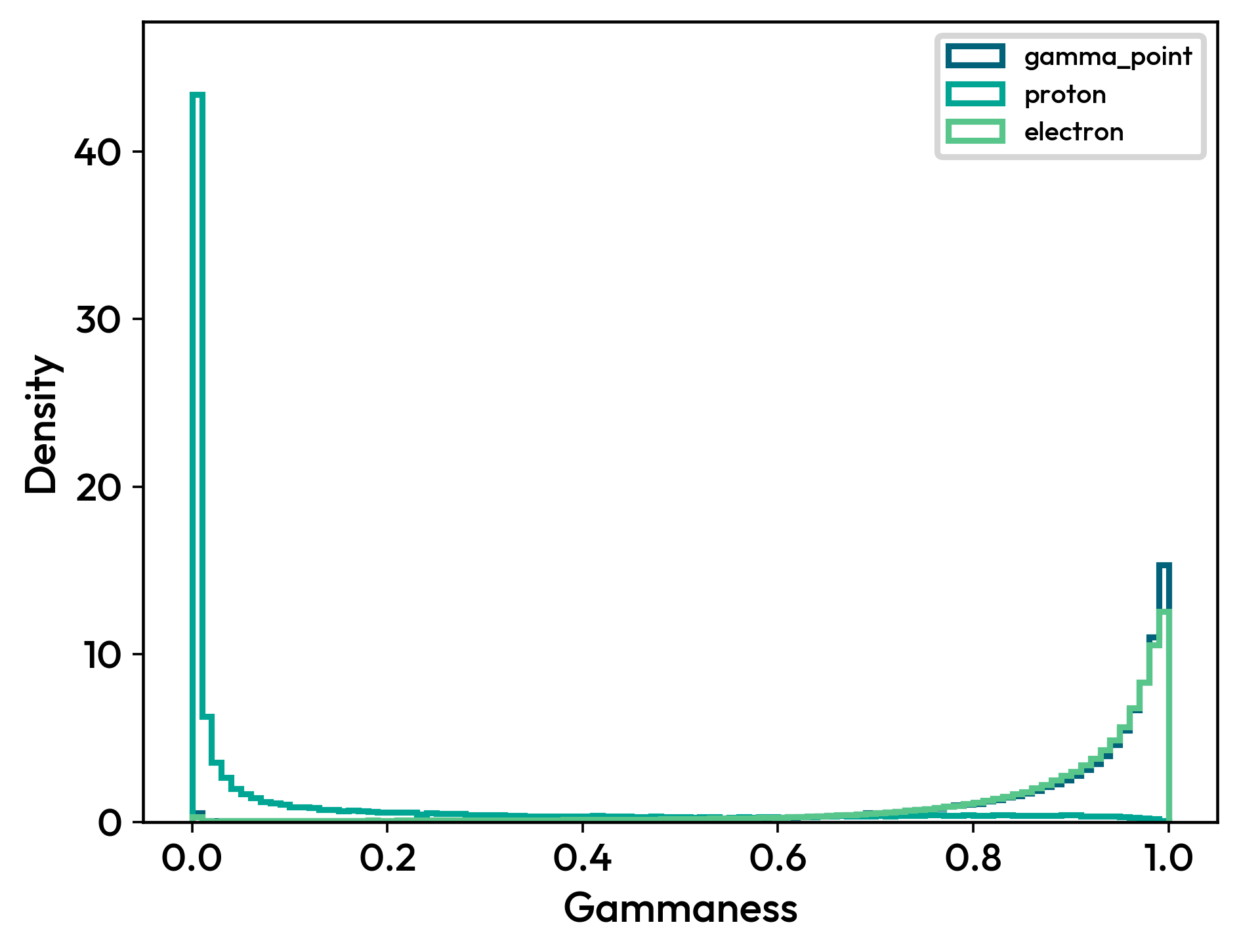} 
        \put(40,50){\textcolor{gray}{Preliminary}}
    \end{overpic}
    \begin{overpic}[width=4.9cm]{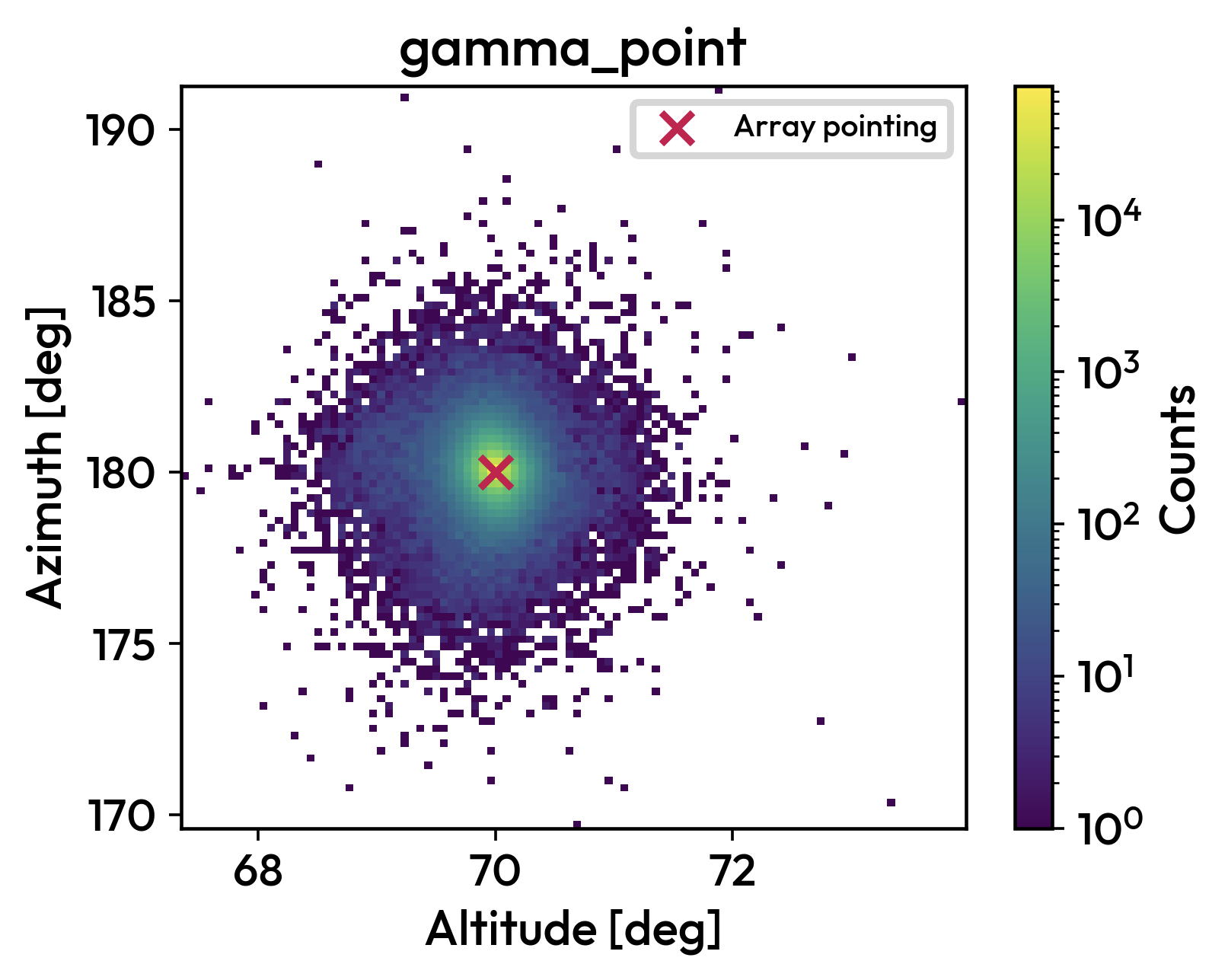} 
        \put(40,18){\textcolor{gray}{Preliminary}}
    \end{overpic}
    \begin{overpic}[width=4.9cm]{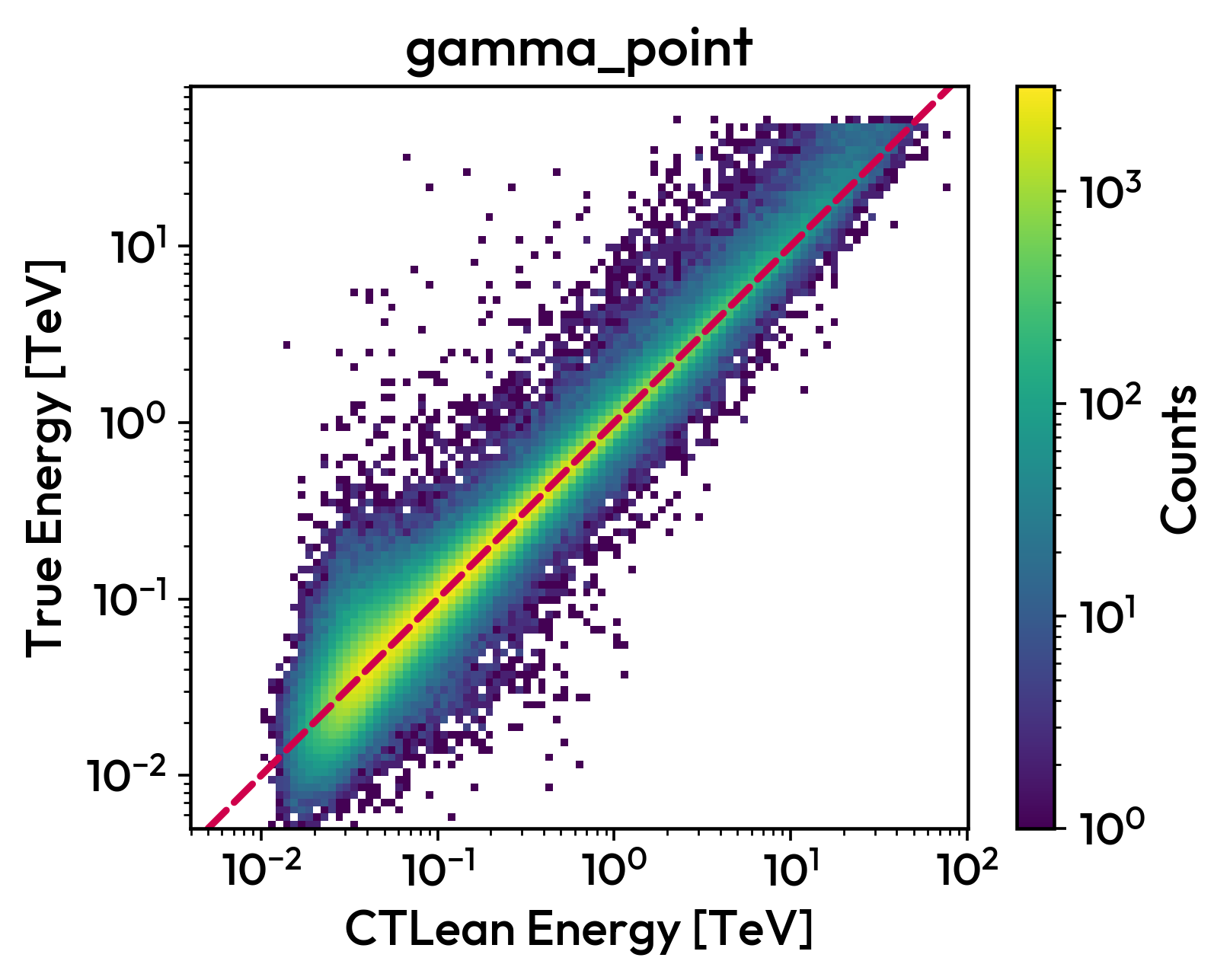} 
        \put(40,18){\textcolor{gray}{Preliminary}}
    \end{overpic}
    \caption{(\textbf{Left}): Distribution of the classifier output (gammaness) for different particle types: electrons, protons, and point-like gamma rays. Higher gammaness values indicate a higher likelihood of being a gamma-ray event, allowing for effective background suppression. (\textbf{Middle}): Reconstructed arrival directions (altitude and azimuth) for point-like gamma-ray events. The true source position corresponds to the center of the field of view, i.e., the array pointing direction. The distribution demonstrates the model’s ability to localize gamma-ray events around the true source location. (\textbf{Right}): Energy migration matrices showing the relationship between true and reconstructed energy for pointlike gamma rays. The color log scale represents the event density in each bin. The red diagonal line indicates perfect energy reconstruction, where the reconstructed energy matches the true energy. Deviations from the line highlight the resolution and possible biases in the energy estimation.}
    \label{fig:particle_classification_altaz_energy}
\end{figure}

Reconstruction performance is validated using simulated test data without applying any cuts on gammaness or direction. The classification model shows clear separation based on the gammaness distribution (see left panel of Fig.~\ref{fig:particle_classification_altaz_energy}). Direction reconstruction accuracy is evaluated by comparing predicted arrival directions to the true source position at the center of the field of view (see middle panel of Fig.~\ref{fig:particle_classification_altaz_energy}). Energy reconstruction is assessed through energy migration matrices, showing good agreement between true and reconstructed energies for pointlike gamma rays (see right panel of Fig.~\ref{fig:particle_classification_altaz_energy}).

\begin{figure}[h]
    \centering
    \begin{overpic}[width=14.9cm]{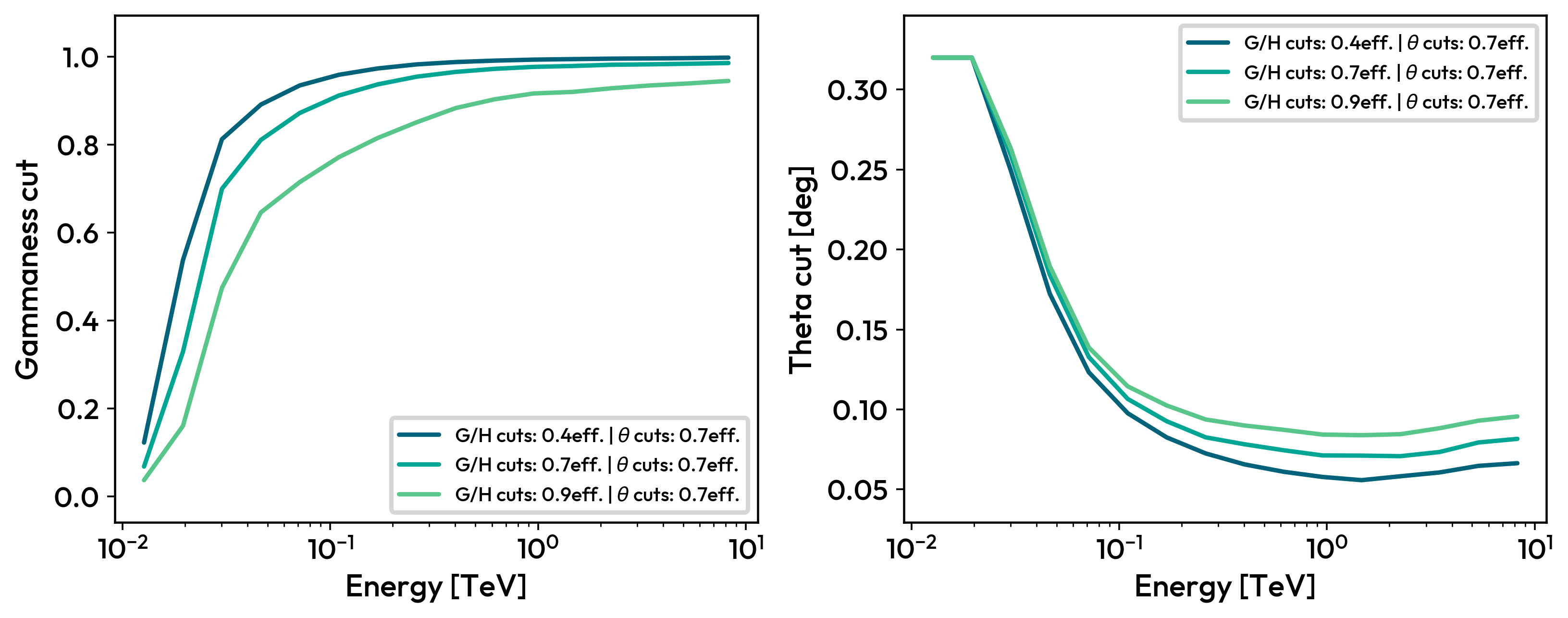} 
        \put(80,20){\textcolor{gray}{Preliminary}}
        \put(30,20){\textcolor{gray}{Preliminary}}
    \end{overpic}
    \caption{Energy-dependent gammaness cut (left) and angular ( $\theta$ ) cut (right) values used for IRF computation values corresponding to three different gamma-ray efficiency levels: 40\%, 70\%, and 90\%. Both energy-dependent cuts are derived as functions of reconstructed energy to optimize performance across the full energy range.}
    \label{fig:cuts}
\end{figure}

\begin{figure}[h]
    \centering
     \begin{overpic}[width=7.4cm]{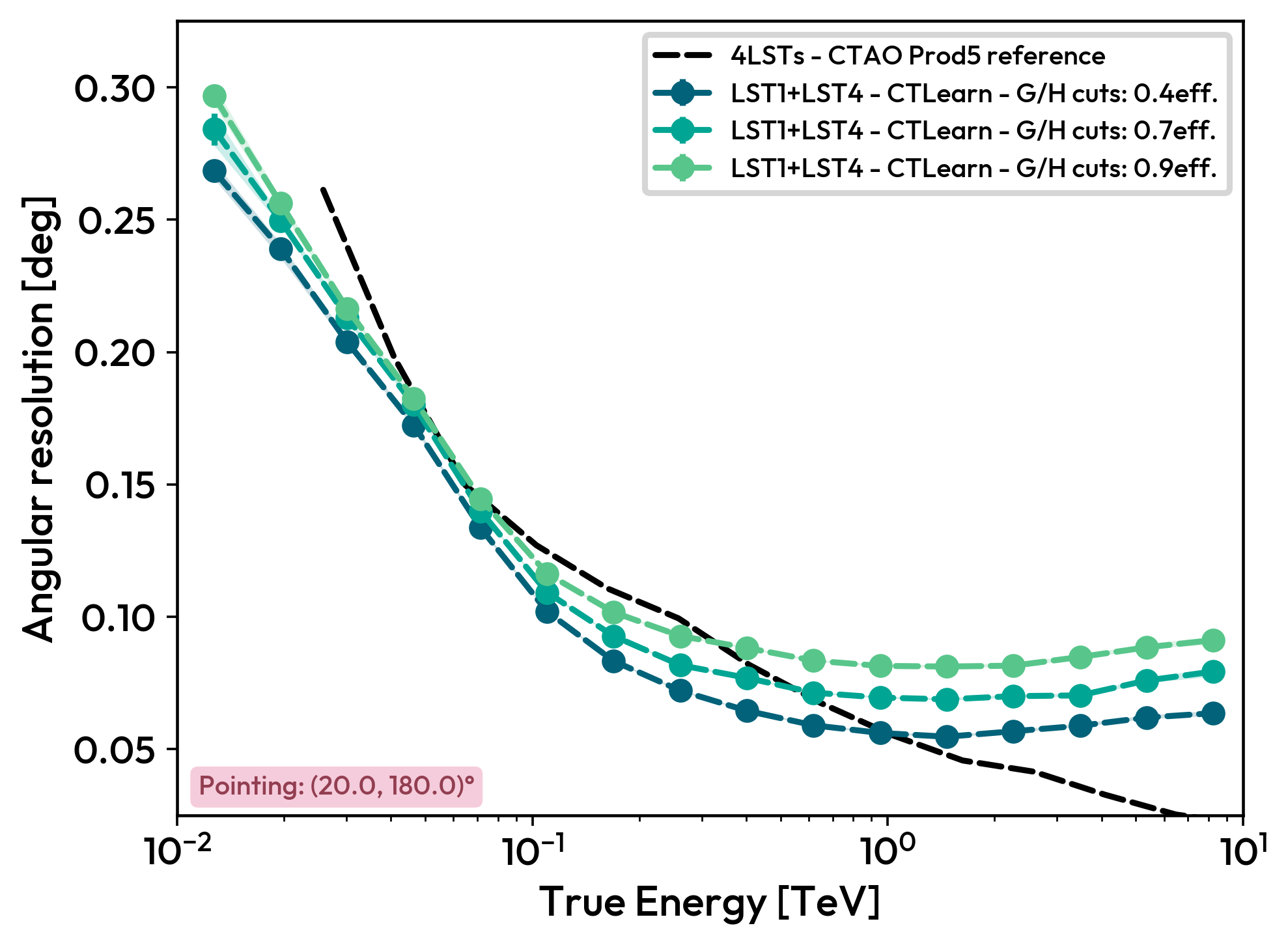} 
        \put(60,50){\textcolor{gray}{Preliminary}}
    \end{overpic}
    \begin{overpic}[width=7.4cm]{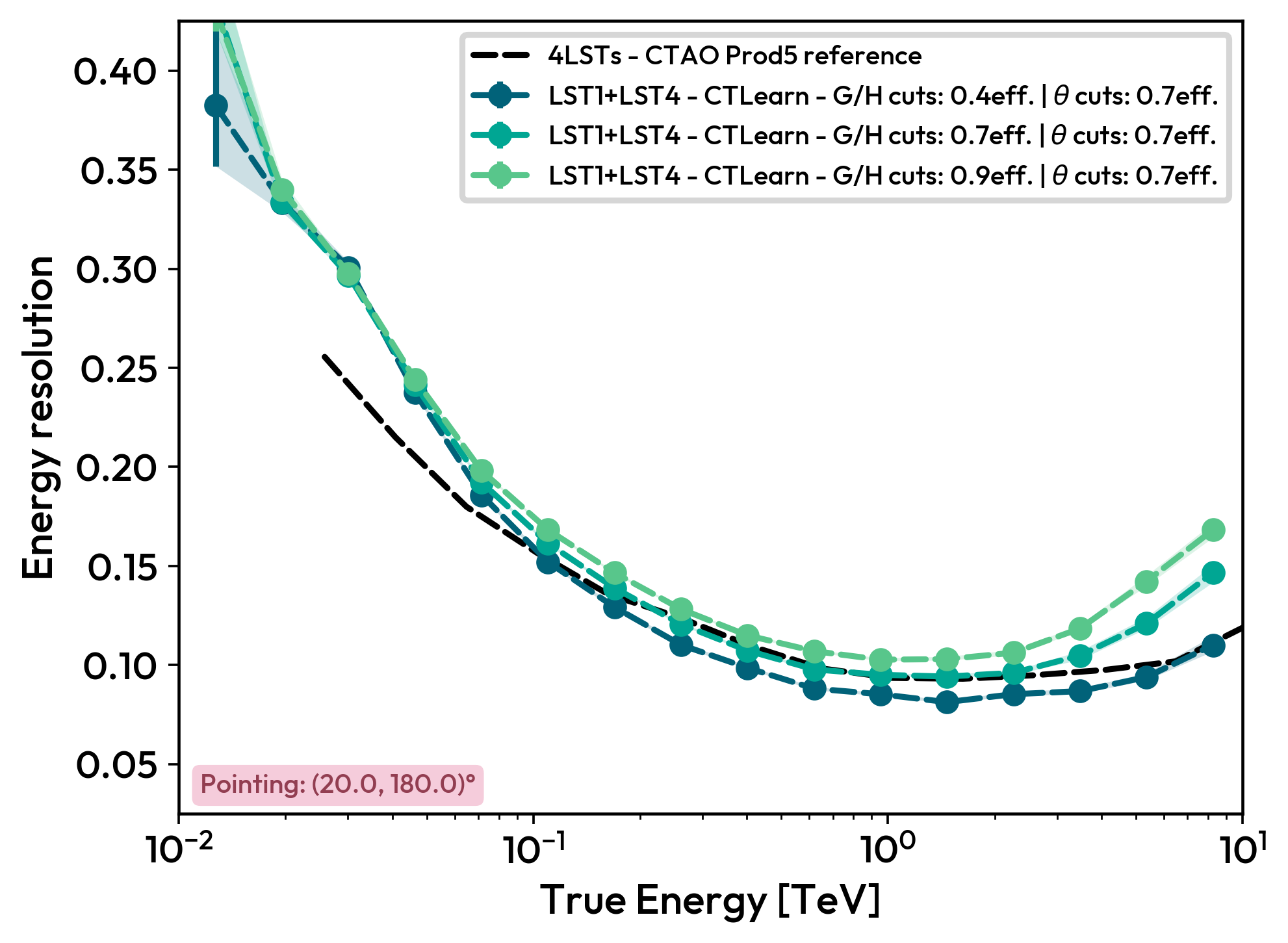}
    \put(60,50){\textcolor{gray}{Preliminary}}
    \end{overpic}
    \caption{Angular (left) and energy (right) resolution as a function of true energy. Both metrics are shown for three different gamma-ray efficiency levels: 40\%, 70\%, and 90\%, illustrating the trade-off between event selection tightness and reconstruction performance. The dashed black line indicates the energy and angular resolution of the LST subarray (4LSTs) from the CTAO \textit{prod5} reference~\cite{cherenkov_telescope_array_observatory_2021_5499840}. It is worth noticing that the shown resolution curves have been obtained exclusively from MC simulation restricted to one particular telescope pointing in the sky.}
    \label{fig:resolutions}
\end{figure}

A global selection requiring an image intensity after cleaning greater than 50 p.e. is applied. This is followed by an energy-dependent cut on gammaness, where a minimum threshold is set in each reconstructed energy bin to retain a fixed fraction of simulated gamma-ray events.  In this study, we used three different retention levels: 40\%, 70\%, and 90\%, to explore the trade-off between background suppression and gamma-ray efficiency. For the computation of the energy resolution (see right panel of Fig.~\ref{fig:resolutions}), an additional energy-dependent angular ($\theta$) cut is applied (see right panel of Fig.~\ref{fig:cuts}). This cut selects the 70\% of gamma-ray events, which survived the gammaness cut, with the most accurate direction reconstruction within each reconstructed energy bin, considering only those events that are well-oriented with respect to the array pointing direction.

 The angular resolution is defined as the angle containing 68\% of the reconstructed gamma-ray events relative to the simulated point source gamma-ray direction. This is calculated in each logarithmic energy bin based on the simulated true energy. The energy resolution in each true energy bin is calculated with 68\% of containment of $ (E_{\mathrm{reco}} - E_{\mathrm{true}})/E_{\mathrm{true}} $.
 
The differential sensitivity calculation per reconstructed energy bin requires a minimal significance of more than 5 $ \sigma $, at least ten detected gamma rays, and a minimal excess over background ratio of 0.05 for observation of 50 hours. The differential sensitivity calculation per energy bin requires a minimal significance of more than 5 $ \sigma $, at least ten detected gamma rays, and a minimal excess over background ratio of 0.05 for observation of 50 hours. The significance of detection is calculated in the IACT community following ``Li\&Ma'' (Eq. 17 in~\cite{LiMa:1983}).

\begin{figure}[h]
    \centering
    \begin{overpic}[width=7.4cm]{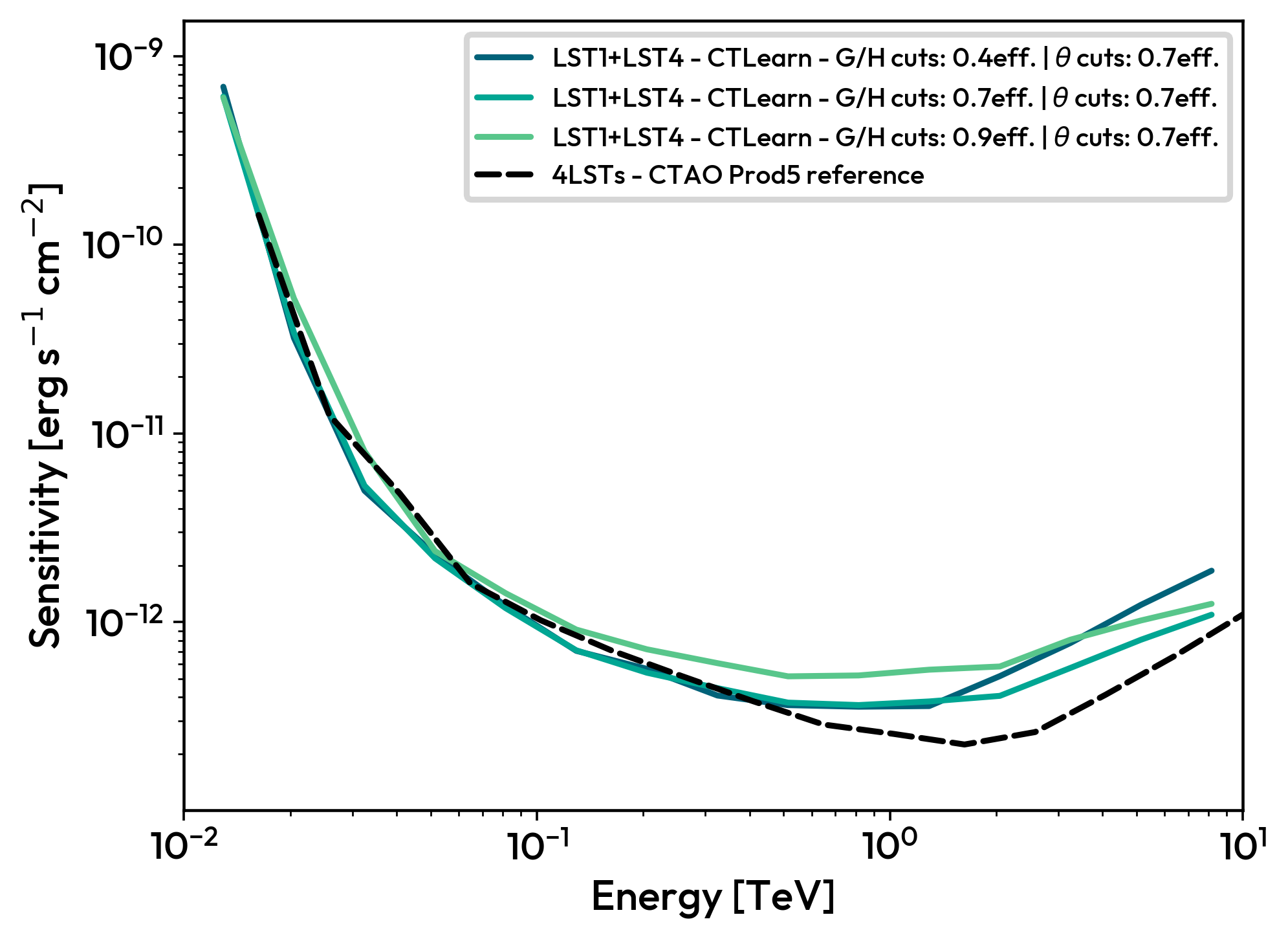}
    \put(60,50){\textcolor{gray}{Preliminary}}
    \end{overpic}
    \caption{Differential sensitivity as a function of reconstructed energy, calculated for three gamma-ray efficiency levels: 40\%, 70\%, and 90\%. The curves illustrate how the choice of event selection efficiency impacts the minimum detectable flux, with tighter cuts improving background rejection at the cost of gamma-ray statistics. The dashed black line indicates the differential sensitivity of the LST subarray (4LSTs) from the CTAO \textit{prod5} reference~\cite{cherenkov_telescope_array_observatory_2021_5499840}. It is worth noticing that the shown sensitivity curves have been obtained exclusively from MC simulation restricted to one particular telescope pointing in the sky.}
    \label{fig:SensitivityCurves}
\end{figure}

\section{Discussions and conclusions}
This study demonstrates the effectiveness of the DL-based \texttt{CTLearn} model for the reconstruction and classification of stereoscopic events recorded by two Large-Sized Telescopes (\texttt{LST1+LST4}) in the CTAO-North array. By leveraging a lightweight TRN architecture and a full DL pipeline for energy, direction, and particle type inference, we achieve competitive reconstruction performance with relatively low computational cost. The stereoscopic configuration enables the model to exploit combined image information, improving the accuracy of event reconstruction and background suppression. The results obtained exclusively from MC simulation validate the capability of the \texttt{CTLearn} framework to produce IRFs and sensitivity estimation for \texttt{LST1+LST4} matching the reference CTAO \textit{prod5} performance for the LST subarray (4LSTs) below $ 1~\text{TeV}$. For the energies below $ 1~\text{TeV}$, the \textit{CTLearn-stackTRN model} for \texttt{LST1+LST4} achieves a better angular resolution than the reference CTAO \textit{prod5} performance of 4LSTs.

This work is especially timely as the construction of the full array of four LSTs at the CTAO-North site is nearing completion. \texttt{LST1+LST4} are entering the stereoscopic commissioning phase, marking a crucial step toward scientific operations. The presented \texttt{CTLearn}-based analysis chain is ready for application to real observational data from these commissioning telescopes. It provides a solid foundation for integrating DL into the early data processing and scientific analysis workflow of CTAO, supporting both performance evaluation and rapid data interpretation during the commissioning and early science phases.

\tiny
\bibliographystyle{unsrturl}
\bibliography{biblio.bib}

\textbf{Full Author List: CTAO-LST Project}

\tiny{\noindent
K.~Abe$^{1}$,
S.~Abe$^{2}$,
A.~Abhishek$^{3}$,
F.~Acero$^{4,5}$,
A.~Aguasca-Cabot$^{6}$,
I.~Agudo$^{7}$,
C.~Alispach$^{8}$,
D.~Ambrosino$^{9}$,
F.~Ambrosino$^{10}$,
L.~A.~Antonelli$^{10}$
% ...
The CTAO-LST Project
}\\

\tiny{\noindent$^{1}${Department of Physics, Tokai University, 4-1-1, Kita-Kaname, Hiratsuka, Kanagawa 259-1292, Japan}.
$^{2}${Institute for Cosmic Ray Research, University of Tokyo, 5-1-5, Kashiwa-no-ha, Kashiwa, Chiba 277-8582, Japan}.
$^{3}${INFN and Università degli Studi di Siena, Dipartimento di Scienze Fisiche, della Terra e dell'Ambiente (DSFTA), Sezione di Fisica, Via Roma 56, 53100 Siena, Italy}.
$^{4}${Université Paris-Saclay, Université Paris Cité, CEA, CNRS, AIM, F-91191 Gif-sur-Yvette Cedex, France}.
$^{5}${FSLAC IRL 2009, CNRS/IAC, La Laguna, Tenerife, Spain}.
$^{6}${Departament de Física Quàntica i Astrofísica, Institut de Ciències del Cosmos, Universitat de Barcelona, IEEC-UB, Martí i Franquès, 1, 08028, Barcelona, Spain}.
$^{7}${Instituto de Astrofísica de Andalucía-CSIC, Glorieta de la Astronomía s/n, 18008, Granada, Spain}.
$^{8}${Department of Astronomy, University of Geneva, Chemin d'Ecogia 16, CH-1290 Versoix, Switzerland}.
$^{9}${INFN Sezione di Napoli, Via Cintia, ed. G, 80126 Napoli, Italy}.
$^{10}${INAF - Osservatorio Astronomico di Roma, Via di Frascati 33, 00040, Monteporzio Catone, Italy}.
% ...
}

\acknowledgments 
\tiny{
IV acknowledges funding for the project “SKYNET: Deep Learning for Astroparticle Physics”, 693 PRIN 2022 (CUP: D53D23002610006). We gratefully acknowledge financial support from the following agencies and organisations:
Conselho Nacional de Desenvolvimento Cient\'{\i}fico e Tecnol\'{o}gico (CNPq), Funda\c{c}\~{a}o de Amparo \`{a} Pesquisa do Estado do Rio de Janeiro (FAPERJ), Funda\c{c}\~{a}o de Amparo \`{a} Pesquisa do Estado de S\~{a}o Paulo (FAPESP), Funda\c{c}\~{a}o de Apoio \`{a} Ci\^encia, Tecnologia e Inova\c{c}\~{a}o do Paran\'a - Funda\c{c}\~{a}o Arauc\'aria, Ministry of Science, Technology, Innovations and Communications (MCTIC), Brasil;
Ministry of Education and Science, National RI Roadmap Project DO1-153/28.08.2018, Bulgaria;
Croatian Science Foundation (HrZZ) Project IP-2022-10-4595, Rudjer Boskovic Institute, University of Osijek, University of Rijeka, University of Split, Faculty of Electrical Engineering, Mechanical Engineering and Naval Architecture, University of Zagreb, Faculty of Electrical Engineering and Computing, Croatia;
Ministry of Education, Youth and Sports, MEYS  LM2023047, EU/MEYS CZ.02.1.01/0.0/0.0/16\_013/0001403, CZ.02.1.01/0.0/0.0/18\_046/0016007, CZ.02.1.01/0.0/0.0/16\_019/0000754, CZ.02.01.01/00/22\_008/0004632 and CZ.02.01.01/00/23\_015/0008197 Czech Republic;
CNRS-IN2P3, the French Programme d’investissements d’avenir and the Enigmass Labex, 
This work has been done thanks to the facilities offered by the Univ. Savoie Mont Blanc - CNRS/IN2P3 MUST computing center, France;
Max Planck Society, German Bundesministerium f{\"u}r Bildung und Forschung (Verbundforschung / ErUM), Deutsche Forschungsgemeinschaft (SFBs 876 and 1491), Germany;
Istituto Nazionale di Astrofisica (INAF), Istituto Nazionale di Fisica Nucleare (INFN), Italian Ministry for University and Research (MUR), and the financial support from the European Union -- Next Generation EU under the project IR0000012 - CTA+ (CUP C53C22000430006), announcement N.3264 on 28/12/2021: ``Rafforzamento e creazione di IR nell’ambito del Piano Nazionale di Ripresa e Resilienza (PNRR)'';
ICRR, University of Tokyo, JSPS, MEXT, Japan;
JST SPRING - JPMJSP2108;
Narodowe Centrum Nauki, grant number 2023/50/A/ST9/00254, Poland;
The Spanish groups acknowledge the Spanish Ministry of Science and Innovation and the Spanish Research State Agency (AEI) through the government budget lines
PGE2022/28.06.000X.711.04,
28.06.000X.411.01 and 28.06.000X.711.04 of PGE 2023, 2024 and 2025,
and grants PID2019-104114RB-C31,  PID2019-107847RB-C44, PID2019-104114RB-C32, PID2019-105510GB-C31, PID2019-104114RB-C33, PID2019-107847RB-C43, PID2019-107847RB-C42, PID2019-107988GB-C22, PID2021-124581OB-I00, PID2021-125331NB-I00, PID2022-136828NB-C41, PID2022-137810NB-C22, PID2022-138172NB-C41, PID2022-138172NB-C42, PID2022-138172NB-C43, PID2022-139117NB-C41, PID2022-139117NB-C42, PID2022-139117NB-C43, PID2022-139117NB-C44, PID2022-136828NB-C42, PDC2023-145839-I00 funded by the Spanish MCIN/AEI/10.13039/501100011033 and “and by ERDF/EU and NextGenerationEU PRTR; the "Centro de Excelencia Severo Ochoa" program through grants no. CEX2019-000920-S, CEX2020-001007-S, CEX2021-001131-S; the "Unidad de Excelencia Mar\'ia de Maeztu" program through grants no. CEX2019-000918-M, CEX2020-001058-M; the "Ram\'on y Cajal" program through grants RYC2021-032991-I  funded by MICIN/AEI/10.13039/501100011033 and the European Union “NextGenerationEU”/PRTR and RYC2020-028639-I; the "Juan de la Cierva-Incorporaci\'on" program through grant no. IJC2019-040315-I and "Juan de la Cierva-formaci\'on"' through grant JDC2022-049705-I. They also acknowledge the "Atracci\'on de Talento" program of Comunidad de Madrid through grant no. 2019-T2/TIC-12900; the project "Tecnolog\'ias avanzadas para la exploraci\'on del universo y sus componentes" (PR47/21 TAU), funded by Comunidad de Madrid, by the Recovery, Transformation and Resilience Plan from the Spanish State, and by NextGenerationEU from the European Union through the Recovery and Resilience Facility; “MAD4SPACE: Desarrollo de tecnolog\'ias habilitadoras para estudios del espacio en la Comunidad de Madrid" (TEC-2024/TEC-182) project funded by Comunidad de Madrid; the La Caixa Banking Foundation, grant no. LCF/BQ/PI21/11830030; Junta de Andaluc\'ia under Plan Complementario de I+D+I (Ref. AST22\_0001) and Plan Andaluz de Investigaci\'on, Desarrollo e Innovaci\'on as research group FQM-322; Project ref. AST22\_00001\_9 with funding from NextGenerationEU funds; the “Ministerio de Ciencia, Innovaci\'on y Universidades”  and its “Plan de Recuperaci\'on, Transformaci\'on y Resiliencia”; “Consejer\'ia de Universidad, Investigaci\'on e Innovaci\'on” of the regional government of Andaluc\'ia and “Consejo Superior de Investigaciones Cient\'ificas”, Grant CNS2023-144504 funded by MICIU/AEI/10.13039/501100011033 and by the European Union NextGenerationEU/PRTR,  the European Union's Recovery and Resilience Facility-Next Generation, in the framework of the General Invitation of the Spanish Government’s public business entity Red.es to participate in talent attraction and retention programmes within Investment 4 of Component 19 of the Recovery, Transformation and Resilience Plan; Junta de Andaluc\'{\i}a under Plan Complementario de I+D+I (Ref. AST22\_00001), Plan Andaluz de Investigaci\'on, Desarrollo e Innovación (Ref. FQM-322). ``Programa Operativo de Crecimiento Inteligente" FEDER 2014-2020 (Ref.~ESFRI-2017-IAC-12), Ministerio de Ciencia e Innovaci\'on, 15\% co-financed by Consejer\'ia de Econom\'ia, Industria, Comercio y Conocimiento del Gobierno de Canarias; the "CERCA" program and the grants 2021SGR00426 and 2021SGR00679, all funded by the Generalitat de Catalunya; and the European Union's NextGenerationEU (PRTR-C17.I1). This research used the computing and storage resources provided by the Port d’Informaci\'o Cient\'ifica (PIC) data center.
State Secretariat for Education, Research and Innovation (SERI) and Swiss National Science Foundation (SNSF), Switzerland;
The research leading to these results has received funding from the European Union's Seventh Framework Programme (FP7/2007-2013) under grant agreements No~262053 and No~317446;
This project is receiving funding from the European Union's Horizon 2020 research and innovation programs under agreement No~676134;
ESCAPE - The European Science Cluster of Astronomy \& Particle Physics ESFRI Research Infrastructures has received funding from the European Union’s Horizon 2020 research and innovation programme under Grant Agreement no. 824064.}

\end{document}